# Validity and reliability of oral temperature compared to ingestible core temperature pill in free-living conditions

Paurakh L. Rajbhandary and Gabriel Nallathambi

Vital Connect Inc., San Jose, CA 95110

Complex thermodynamics of the human body and external factors may cause differences between oral and core body temperatures. We present a 19-subject study where continuous measurements of ingested core temperature pills are compared against spot measurements of oral temperature, both FDA-cleared devices, in free-living conditions. Based on measurements from 419 samples across 19 subjects, MAE of 0.30 ± 0.26 ºC and 95% limits of agreement of -0.60 and 0.90 ºC was observed between ingestible pill temperature and oral temperature. These results demonstrate that statistically significant differences (p<0.001) between oral and core temperature may arise in free-living conditions. Such clinically relevant differences in temperature measurements due to complex thermodynamics of heat exchange in the human body, measurement noise, and other external factors show the intricacies of interpreting body temperatures measured at different locations.

## 1. INTRODUCTION

Continuous monitoring of core temperature as measured at the deep tissues of the body using invasive devices such as pulmonary arterial or esophageal catheter, urinary catheter, and rectal probe is part of routine clinical practice in critical care settings. While vital signs such as heart rate, respiration rate, and oxygen saturation can be continuously monitored using accurate non-invasive devices, clinicians predominantly rely upon spot measurements of less accurate non-invasive oral thermometers in the general wards and out-patient settings to approximate core body temperature due to ease of use [1, 2].

Recent developments have demonstrated the possibility of monitoring core body temperature continuously using less invasive ingestible, wireless sensors [3]. Prior research has established that the core temperature measured by the ingestible sensors is at least as accurate as more invasive rectal probe temperature monitors [4]. However, research comparing non-invasive temperature measurements with core temperature measured by the ingestible sensors is very scarce and limited to controlled exercise trials in an environmental chamber [5]. The goal of this paper is to examine the validity and reliability of clinical grade oral thermometer compared to ingestible temperature pill in free-living conditions typical of routine use and provide insights into potential implications of differences of these temperatures in clinical context.

## 2. MATERIALS AND METHODS

### 2.1. Experimental Protocol

Nineteen subjects participated in an independent IRB approved protocol for three days comprising circadian cycles of core temperature, daily activities, sleep cycles, clothing and environmental effects that may cause variations in body temperature naturally. Each participant received a Welch Allyn SureTemp® Plus 690 Electronic Thermometer [6], FDA cleared VitalSense core temperature pill [3] that communicates and transmits data to a Sensor Electronic Module (SEM) housed in the Equivital life monitor belt [7] with external battery pack, and oral temperature spreadsheet for recording the manual measurements.

The study coordinator first activates the temperature capsule and pairs it with the SEM monitor. After confirming that the SEM monitor was receiving temperature data from the capsule, the participant swallows the core temperature pill with a full glass of water. The SEM monitor was then connected to the Equivital sensor belt such that it was within 1 meter of the subject at all times. Subjects were then required to follow their daily routines and activities while wearing the Equivital sensor belt. Participants were instructed to wear the Equivital sensor belt at all times, except when changing or showering. In case the core body temperature pill was determined to be out of the participant's body, the participant was instructed to continue the protocol for up to 72 hours.

Participants were trained to take oral measurements per standard clinical procedure. Participants manually recorded all oral measurements and time of measurement on the spreadsheet provided by the study coordinator. Ten to sixteen oral temperature measurements (3 readings at each timepoint) are recommended throughout the day covering immediately after the participant wakes up before falling asleep and during daily routines. Participants were instructed to take oral measurements at least 30 minutes after consumption of food or liquid, showering or sun exposure. Finally, participants were asked to avoid strenuous or extensive physical activity during the 72 hours of protocol.

## 2.2. Data analysis

The core temperature pill continuously transmitted temperature data every 15 seconds. For data analysis purposes, the average of the three recorded oral measurements at each timepoint and the core temperature pill data nearest to the spot measurements of oral temperature in a 15 minutes two-sided window was selected. All data collected from the pill within 2 hours before and after the ingestion and excretion were excluded.

The accuracy of the measurements was established in terms of mean absolute error (MAE) and bias (b) calculated as follows:

$$b_{average\ subject} = \frac{1}{n}\sum_{j=1}^{n}\frac{1}{N_j}\sum_{i=1}^{N_j}\left(t_{core_{i,j}} - t_{oral_{i,j}}\right)$$

$$MAE_{average\ subject} = \frac{1}{n}\sum_{j=1}^{n}\frac{1}{N_j}\sum_{i=1}^{N_j}\left|t_{core_{i,j}} - t_{oral_{i,j}}\right|$$

$$b_{gross} = \frac{1}{N_j\ n}\sum_{j=1}^{n}\sum_{i=1}^{N_j}\left(t_{core_{i,j}} - t_{oral_{i,j}}\right)$$

$$MAE_{gross} = \frac{1}{N_j\ n}\sum_{j=1}^{n}\sum_{i=1}^{N_j}\left|t_{core_{i,j}} - t_{oral_{i,j}}\right|$$

In the above equations, $n$ is the total number of subjects, $N_j$ is the total number of samples used for comparison for the $j$-th subject, $t_{core_{i,j}}$ is the $i$-th ingestible core pill temperature measurement of $j$-th subject, and $t_{oral_{i,j}}$ is the $i$-th oral temperature measurement of $j$-th subject. Metrics are calculated in terms of both average and gross statistics.

## 3. RESULTS AND DISCUSSION

Figure 1 shows an example of a subject's ingestible pill temperature along with spot measurements of oral temperature across multiple days in free-living conditions except strenuous exercise. The oral temperature readings are consistently lower than the ingestible pill temperature. While the trend information is often lost due to sporadic measurements of oral temperature, the ingestible pill temperature continuously tracks the rise and fall in temperature seamlessly.

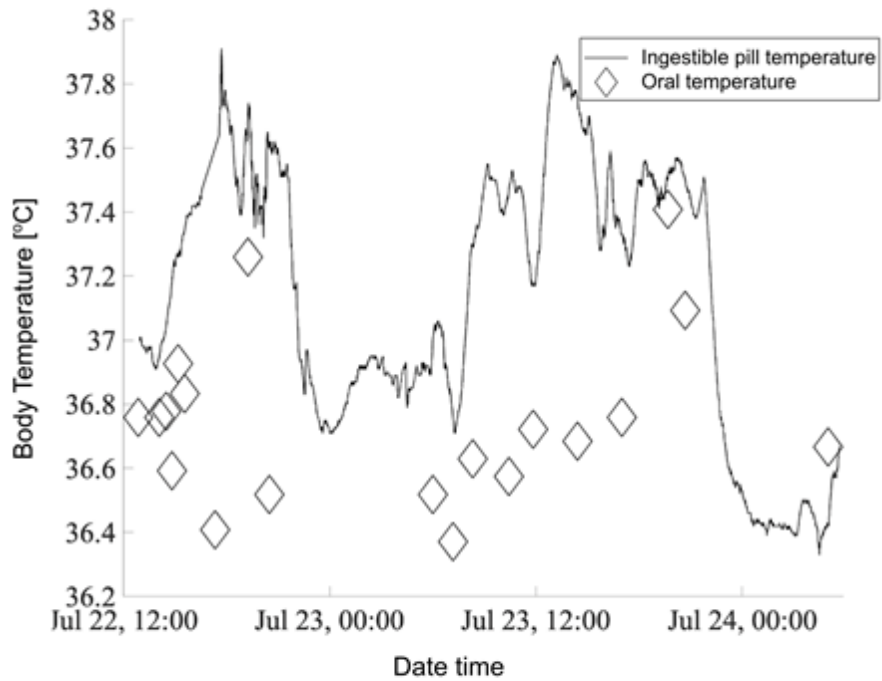

*Figure 1: VitalSense core body temperature compared to Welch Allyn SureTemp 690 oral temperature time series for a subject.*

The results are summarized in Table I and the Bland-Altman plot is provided in Figure 2. MAE and bias based on gross statistics were 0.30 ± 0.26 ºC and 0.14 ± 0.37 ºC based on 419 numbers of samples with 95% limit of agreement of -0.6 and 0.9 ºC (p<0.001 based on Wilcoxon signed rank test). Mean and standard deviation of MAE and mean of bias calculated per subject was 0.31 ± 0.13 ºC and 0.18 ± 0.21 ºC, respectively. Mean duration of data collected from the pill with 2 hours before and after the ingestion and excretion excluded was 38.4 ± 16.4 hours.

Table I: Performance of oral temperature compared to ingestible core temperature pill (mean ± standard deviation).

| Parameter | Average statistics | Gross statistics |
|---|---|---|
| N [ Subjects or Samples] | 19 | 419 |
| MAE [ºC] | 0.31 ± 0.13 | 0.30 ± 0.26 |
| Bias [ºC] | 0.18 ± 0.21 | 0.14 ± 0.37 |
| Ingestion data duration [hours] | 38.4 ± 16.4 | 729.6 |

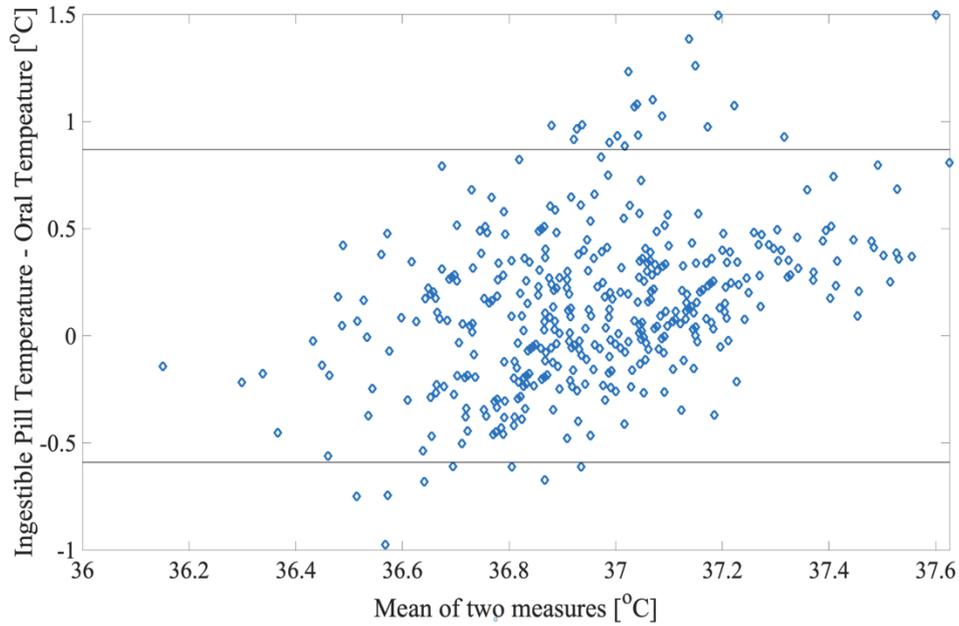

*Figure 2: Bland-Altman plot for core and oral temperature measurements in free-living conditions.*

While a previous study found a positive bias of 0.6 ºC ($p<0.001$) for ingestible pill temperature over oral temperature with 95% limit of agreement of -1.9 and 1.9 ºC during exercise in heat [2], the present study shows that a statistically significant difference exists also in free-living conditions. These findings suggest that oral temperature may not provide an exact measurement of core temperature.

As a separate test, the two aforementioned temperature measurement devices were also compared during a strenuous indoor treadmill run for approximately two hours following resting under the blanket without falling asleep for approximately 3 hours (Figure 3). Oral temperatures were recorded periodically by allowing the subject to rest for 2 minutes at a time. Oral measurements were taken at least five minutes after ingesting any water or fluid orally. During the period when the subject is laying down under the blanket, ingestible pill temperature gradually rose from approximately 37 ºC to 37.6 ºC. The temperature difference between the ingestible pill and the oral temperature slightly increased during this duration. This may be attributed partly to the fact that the blanket covered all parts of the body from neck down while the head of the subject was at room temperature. During the segment when the subject was running on the treadmill, there is a significant difference between the ingestible pill and the oral temperature compared to the previous segment where the subject is lying down under the blanket. At its peak, the ingestible pill recorded a core temperature value that was approximately 2 °C higher than oral measurements. These results are consistent with a prior study that compared ingestible pills to oral temperature during controlled exercise in heat [5].

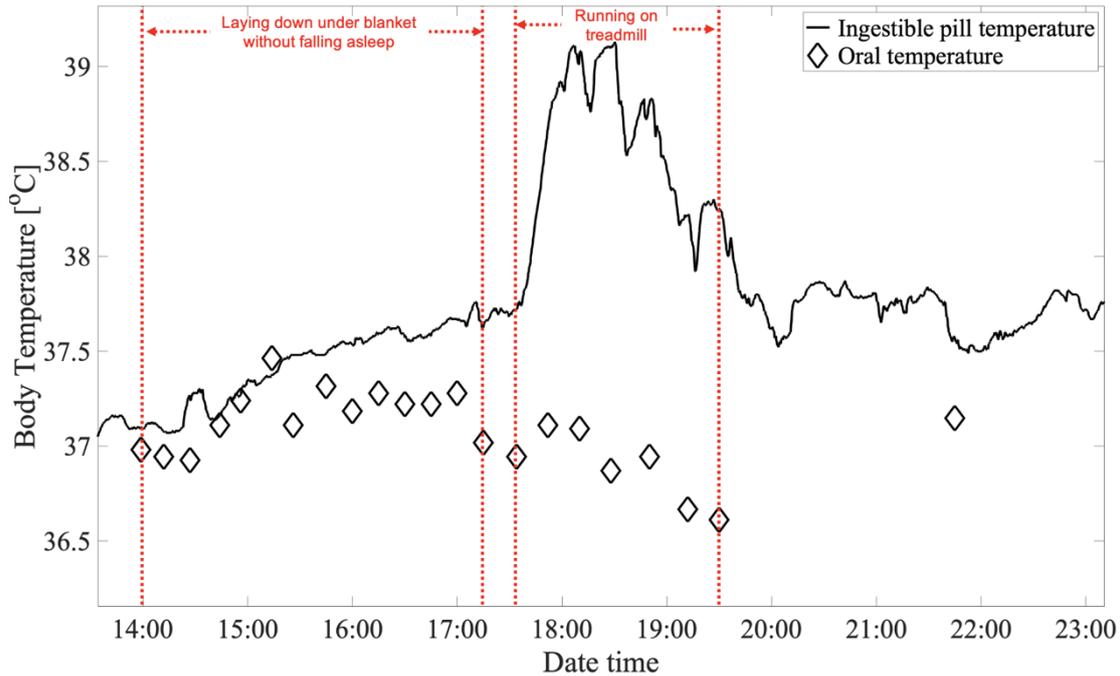

*Figure 3: A separate test where the subject was laying down under a blanket without falling asleep followed by approximately 2 hours of strenuous exercise (running) on the treadmill. Each oral temperature shown is an average of 3 independent measurements.*

Per current clinical practice, non-invasive measurement of temperature from different locations such as oral, tympanic, axillary, etc. serves as a surrogate of core temperature. In Figure 4, based on prior research works [8, 9] and this study, the different body temperature measurements in free-living conditions are compared in terms of accuracy, frequency, usability, and degree of invasiveness.

Statistically significant and potentially clinically relevant differences between the surrogate measures of temperature and core temperature can occur due to differences in offset or bias, phase delay, complex thermodynamics of heat exchange in human body, measurement noise, and environmental factors such as ingestion of beverage, convection of surrounding air, and ambient temperature. These differences in estimate of the measured temperature from the core temperature also depend upon physiological (e.g., normal and febrile body state) and physical changes (e.g., strenuous exercise). The results of this study demonstrate that statistically significant ($p<0.001$) and clinically relevant differences between oral temperature and core temperature exist even during free-living conditions.

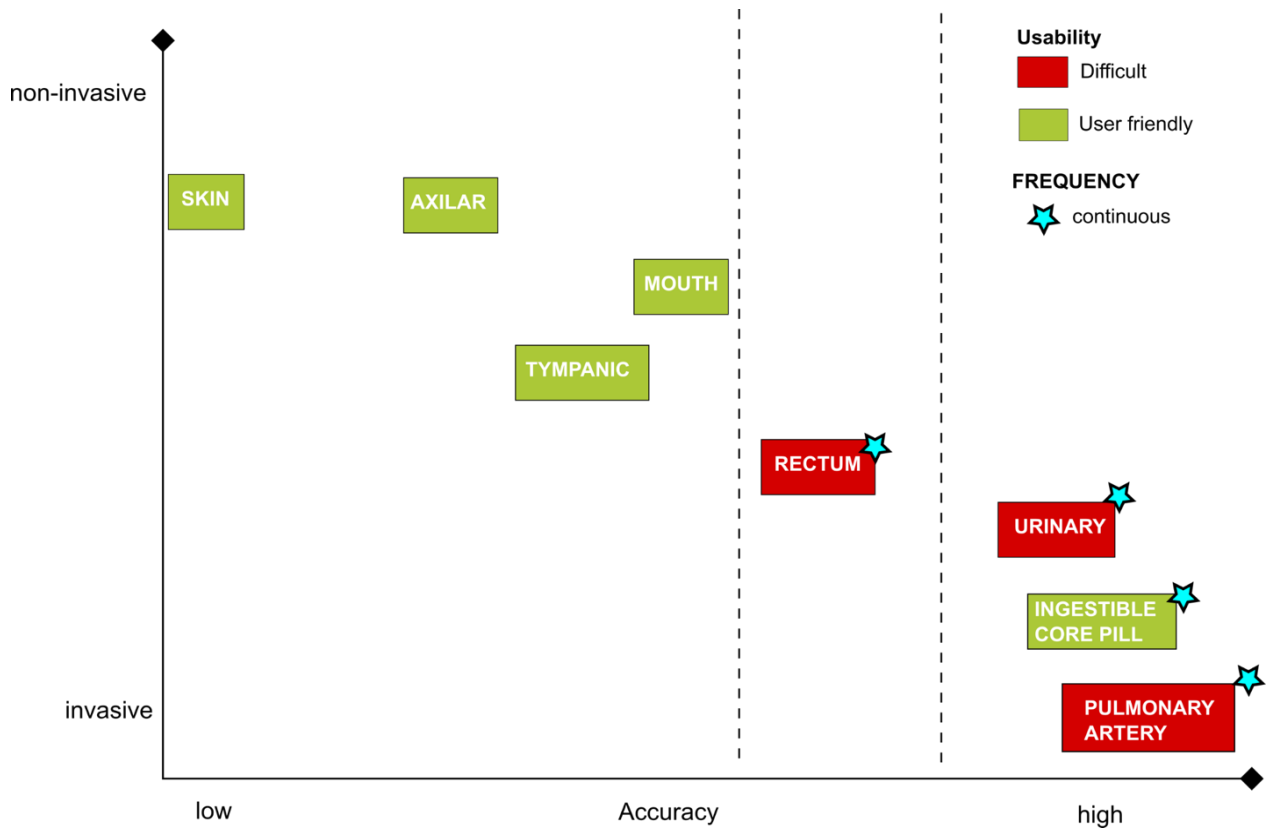

*Figure 4: Comparison of different body temperature measurements in free-living conditions in terms of accuracy, frequency, usability, and degree of invasiveness [8, 9]*

In summary, our findings suggest that oral temperature may not be an accurate and reliable measurement of core temperature. Clinical interpretation should consider all relevant differences prior to approximating core body temperature by non-invasive measures of temperature from different locations of the body.

## References

[1] Evans, J., & Kenkre, J. (2006). Current practice and knowledge of nurses regarding patient temperature measurement. Journal of medical engineering & technology, 30(4), 218-223.

[2] Ilsley, A. H., Rutten, A. J., & Runciman, W. B. (1983). An evaluation of body temperature measurement. Anaesthesia and intensive care, 11(1), 31–39.

[3] McKenzie, J. E., & Osgood, D. W. (2004). Validation of a new telemetric core temperature monitor. Journal of Thermal Biology, 29(7-8), 605-611.

[4] Sparling, P. B., Snow, T. K., & Millard-Stafford, M. L. (1993). Monitoring core temperature during exercise: ingestible sensor vs. rectal thermistor. Aviation, space, and environmental medicine, 64(8), 760-763.

[5] Fogt, D. L., Henning, A. L., Venable, A. S., & McFarlin, B. K. (2017). Non-invasive measures of core temperature versus ingestible thermistor during exercise in the heat. International journal of exercise science, 10(2), 225.ss

[6] Giuliano, K. K., Giuliano, A. J., Scott, S. S., MacLachlan, E., Pysznik, E., Elliot, S., & Woytowicz, D. (2000). Temperature measurement in critically ill adults: a comparison of tympanic and oral methods. American Journal of Critical Care, 9(4), 254-261.

[7] Liu, Y., Zhu, S. H., Wang, G. H., Ye, F., & Li, P. Z. (2013). Validity and reliability of multiparameter physiological measurements recorded by the Equivital Life Monitor during activities of various intensities. Journal of occupational and environmental hygiene, 10(2), 78-85.

[8] Quast, S., & Kimberger, O. (2014). The Significance of Core Temperature—Pathophysiology and Measurement Methods. Dräger Medical GmbH: Lübeck, Germany.

[9] Shahrooz, M. (2017). Re-inventing Core Body Temperature Measurement. Master of Science Thesis, KTH Industrial Engineering and Management.